\documentclass[11pt]{article}
\newif\ifpdf
\ifx\pdfoutput\undefined
\pdffalse % we are not running PDFLaTeX
\else
\pdfoutput=1 % we are running PDFLaTeX
\pdftrue
\fi

\ifpdf
   \usepackage[pdftex,urlcolor=blue,colorlinks=true]{hyperref}
   \usepackage[pdftex]{graphicx}
\else
   \usepackage[ps2pdf,urlcolor=blue,colorlinks=true]{hyperref}
   \usepackage{graphicx}
\fi

\newcommand{\be}{\begin{equation}}
\newcommand{\ee}{\end{equation}}
\newcommand{\bea}{\begin{eqnarray}}
\newcommand{\eea}{\end{eqnarray}}
\begin{document}

\ifpdf
\DeclareGraphicsExtensions{.pdf, .jpg, .tif}
\else
%\DeclareGraphicsRule{.jpg}{eps}{.eps}{`convert #1 eps:-}
\DeclareGraphicsExtensions{.eps, .jpg}
\fi

\title{Angular momentum localization in oval billiards    
}
\author{
    Jens U.~N{\"o}ckel$^*$\\  
Nanovation Technologies\\
Evanston, IL 60201, USA
    }
\date{$^*${\footnotesize
    \href{http://darkwing.uoregon.edu/~noeckel}{\em Current
      address}:\\
Department of Physics,
    University of Oregon, 1371 E 13th Avenue, Eugene, OR 97403\\
Received August 7, 2000; published in Physica Scripta {\bf T
    90}, 263 (2001)}}
\maketitle
PACS: 05.45.Mt, 42.15.-i, 42.25.-p
\begin{abstract}
    Angular momentum ceases to be the preferred basis for identifying 
    dynamical localization in an oval billiard at large excentricity. 
    We give reasons for this, and comment on the classical 
    phase-space structure that is encoded in the wave functions 
    of ``leaky'' dielectric resonators with oval cross section. 
\end{abstract}

Geometric optics is an important engineering tool because of its 
explanatory and predictive power, even when wave effects are present, 
as is the case in resonant cavities. Nevertheless, 
quantum chaos has not been widely recognized as an issue 
in optical resonators until recently \cite{stone1}, because the engineer 
often has the freedom to choose geometries for which either the ray 
picture is simple or the wave equation is separable (up to small
perturbations). 
This is a luxury that we do not usually have in naturally occuring, 
``self-assembled'' optical resonators such as, e.g., aerosol 
droplets \cite{mekis,sschang} or microcrystallites \cite{laeriprl}.; 
these examples typically have mixed phase spaces. 
What we learn from such systems in turn allows us 
to accept chaotic ray dynamics as a way to introduce added freedom 
into the design of optical devices in a wide range of material systems, 
such as semiconductor microdisks \cite{slusher,ho}, polymers and glasses
\cite{dodabalpur,Haroche,arnold}. One of the essential phenomena that 
makes chaotic resonators useful in this respect is dynamical 
localization, because it allows cavity resonances with decay 
rates that exceed the values expected from classical ray 
considerations. Mixed dynamics does not necessarily make it 
impossible to identify localization \cite{raizen}, provided the 
classical phase space stucture is properly taken into account. In this 
paper, we discuss how localization can be characterized in 
oval dielectric cavities. 

From the quantum-chaos perspective, dielectric microcavities allow us 
to study the ray-wave duality in a class of open billiard systems bounded 
by ``penetrable'' walls which introduce an escape condition in phase 
space \cite{borgonovi1}. 
This openness arises because the internal and external region 
are coupled across the dielectric-air interface. In many cases this 
interface can be considred as abrupt on the scale of the wavelength, 
in which case one arrives at a set of polarization-dependent 
dielectric boundary conditions which in the ray limit correspond to 
Fresnel's laws of reflection. The latter have two basic 
consequences: (a) if the cavity has refractive index $n$ and the 
outside is assumed to be air, then rays hitting the surface with 
angle of incidence greater than $\chi_{c} \equiv \arcsin 1/n$, 
measured with respect 
to the normal, undergo total internal reflection. (b) the interface 
exhibits a finite reflectivity even at normal incidence, $\chi=0$, 
given by $R=(n-1)^2/(n+1)^2$; this ``ray splitting'' implies that 
rays violating the total-internal-reflection condition 
may still continue along an internal trajectory with attenuated 
amplitude \cite{kohler}. In fact, for large refractive index $n$, the 
limit of a closed cavity with reflectivity $R=1$ is approached. 

In the quantum-classical transition under such 
circumstances, the competition between the internal time scales (as 
set most prominently by the density of levels) and 
the state-dependent decay rates must be taken into 
account \cite{prigogine,haenggi,braun}. This becomes especially 
interesting in cavities with mixed phase spaces because of their 
intricate temporal evolution \cite{zaslavsky,ketzmerick}. The main 
objects of study in microlaser design are {\em single, isolated} 
resonances. The reason is that the properties of a laser are 
typically determined by the spatial and emission characteristics of 
only one or a few quasibound states. In a single-mode laser, it is the 
state whose $k$ lies closest to the real axis \cite{beenakker2}. 
In contrast to the random-wave assumption that is justified in the 
presence of hard chaos \cite{eckhardt2}, highly anisotropic intensity 
patterns of wave functions are typical for mixed systems. These are 
in fact desirable in a laser because anisotropy can translate to 
{\em focused} emission\cite{optlett1}. 
Individual quasibound states can be studied in great detail in 
microlaser experiments, because one can make spatially and spectrally 
resolved images of the emitter under various observation angles 
\cite{sschang}. 

%\subsection{Formulation of the wave problem}\label{sec:problem}
The numerical aspects of the electromagnetic scattering problem are 
challenging and have several decades of history, particularly in 
atmospheric 
sciences. If the dielectric constant can be assumed piecewise constant 
in the spatial domains defining the scatterer, one computational 
method is that of wavefunction matching: in each dielectric region, a 
``Treftz basis'' is introduced \cite{treftz}, consisting of free-space 
stationary solutions at a fixed wavenumber $k$. The unknown expansion 
coefficients of a true wave solution in this basis are determined by 
imposing the dielectric boundary conditions. In the present study, we 
are interested in quasibound states of a {\em cylindrical} dielectric 
surrounded by air. For simplicity, we specialize to the case where the 
electric field is polarized parallel to the cylinder axis, so that 
Maxwell's equations reduce to a scalar wave equation \cite{mcbook}. 
The internal and external wave functions
\bea\label{psiintdecompgeneqn}
\psi_{\rm int}(r,\phi)&=&\sum\limits_m A_m\,J_m(nkr)\,e^{im\,\phi},\\
\psi_{\rm ext}(r,\phi)&=&\sum\limits_m B_m\,H_m^{(1)}(kr)\,e^{im\,\phi}.
\label{psiintdecompgeneqn1}
\eea
satisfy the usual quantum mechanical boundary conditions at the 
interface,
\be\label{eq:bconds}
\psi_{\rm int}=\psi_{\rm ext}\quad{\rm and}\quad
\partial\psi_{\rm int}/\partial{\bf e}=\partial\psi_{\rm 
ext}/\partial{\bf e}.
\ee
Here, $r$ and $\phi$ are polar coordinates, ${\bf e}$ the outward 
normal; $k$ is the 
free-space wavenumber, and $m$ labels angular momentum. Implicit in 
the form of $\psi_{\rm ext}$ is the condition that only outgoing 
cylindrical waves be present in the exterior, i.e. the solutions will 
represent emission without any incoming wave \cite{thesis} -- 
appropriate for fluorescence or lasing. The resulting 
homogenous system of Eq.\ (\ref{eq:bconds}) for the $A_{m}$, $B_{m}$ 
has a nontrivial solution only at a set of discrete, complex values 
of $k$. 

One advantage of Eq.\ (\ref{psiintdecompgeneqn}) is that it permits a 
reformulation in terms of the internal scattering matrix of the 2D 
cross-sectional billiard, in the sense of the ``scattering approach to 
quantization'' \cite{smilansky}. However, it is by no means clear 
that these expansions in angular momentum eigenfunctions with 
coefficients $A_{m}$, $B_{m}$ converge. The 
assumption that this is in fact the case 
is known as the Rayleigh hypothesis \cite{vdberg,barton}. 
For definiteness, the convex boundaries which 
shall serve as our model systems are parametrized in polar 
coordinates by two-dimensional multipoles of constant area,
\be\label{eq:shapedefin}
r(\phi)= R\,\left(1 + 
\epsilon\cos(\nu\phi)\right)/\sqrt{1+\epsilon^2/2},
\ee
where $\nu=1,\,2\ldots$ and $\epsilon$ measures the fractional 
deformation. The simplest cases are the Limacon shape ($\nu=1$) and 
the quadrupole ($\nu=2$). Convergence problems arise if the cross 
section is too strongly deformed, so that the radii of convergence 
of the inner or outer expansion, Eqs.\ 
(\ref{psiintdecompgeneqn}) and (\ref{psiintdecompgeneqn1})  
intersect the boundary; this can make it impossible to formulate the 
matching conditions. 

As long as the shape is convex, however, numerical experience 
shows \cite{thesis} that the problem can be regularized over a wide 
range of deformations by performing the wavefunction matching at a 
discrete number $N$ of points in real space and making the 
number $M$ of angular momenta $m$ in Eq.\ (\ref{psiintdecompgeneqn}) 
smaller than $N$. The resulting rectangular matrix problem can then be 
solved by singular-value decomposition \cite{penrose}. Additional 
improvment can be achieved by finding two or more choices of origin 
for the polar coordinates such that the respective domains of convergence 
for all resulting versions of Eq.\ (\ref{psiintdecompgeneqn}), 
taken together, cover the boundary completely. The additional 
unknowns in these expansions are then connected by {\em analytical 
continuation}. A simple example for this analytical-continuation 
approach is the annular billiard \cite{doron,hackenbroich}, in which 
a circle with an eccentric circular inclusion permits expansions of 
the type Eq.\ (\ref{psiintdecompgeneqn}), centered either at the inner 
or outer circle; the connection between the two expansions is given 
analytically via the addition theorems for Bessel functions.

There are no truly 
bound states in this finite-sized 2D photonic system; the same is true 
for the generalization to a three-dimensionally confined cavity of 
finite extent. This 
is the reason for having to permit complex $k$ above, assuming $n$ is 
real. The 
fact that all states are metastable distinguishes these systems from 
the otherwise similar subject of attractive wells in quantum mechanics.
This is a reminder of the {\em inequivalence} between optical and 
quantum-mechanical wave equations. Microwave experiments 
\cite{stein, arichter, sridhar} can under certain restrictions 
emulate of Schr\"odinger's equation, if the genuinely electrodynamic 
aspects of the resonator (resulting from the vectorial nature and 
hyperbolic charactistics of Maxwell's time-dependent wave equations) 
are not 
important. In dielectric optical cavities, on the other hand, one is 
often forced by the intended applications to go beyond this 
analogy \cite{amsterdam}. 

%\subsection{Relation between the open system and dissipation}

%\section{Analytical approximation for the classical billiard map}
\begin{figure}[t]
\centering{
\includegraphics[width=6cm]{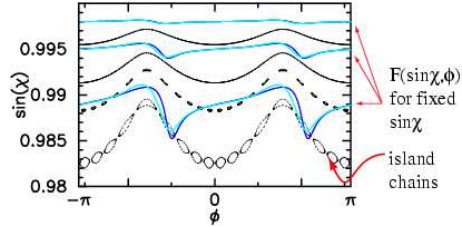}}
        \caption{A narrow strip of the Poincar{\'e} section for a 
        quadrupole with $\epsilon=0.1$, magnifying the 
        whispering-gallery region. Superimposed on the high-order 
        island chains and invariant KAM curves are plots of the kick strength 
        functions.
        }
        \label{fig:kickstrength}
    \end{figure}
As was first argued in Ref.\ \cite{optlett1}, the 
Poincar{\'e} map of the billiard, Eq.\ (\ref{eq:shapedefin}), 
contains the essential information determining the emission 
characteristics of the corresponding resonator. Therefore, it is 
necessaery to understand the properties of this 
map in detail. It can be written in the form 
\bea\label{eq:sinchimap}
\sin{\bar\chi}&=&\sin\chi+F({\bar s},\sin\chi),\\
{\bar s}&=&s + G(s,\sin\chi),\label{eq:smap}
\eea
where $s$ is the arc length along the boundary and the bar denotes the 
new coordinates after one iteration of the map. 
The functions $F$ and $G$ contain the nonlinearity of the dynamics, 
as shown in Fig.\ \ref{fig:kickstrength} where we plot $F$ versus 
final position $s$ for three fixed values of $\sin\chi$. Note that the 
amplitude of the nonlinearity goes to zero as $\sin\chi\to 1$. To 
quantify this, Fig.\ \ref{fig:kickaverage} plots the 
root-mean-square of $F$ versus starting 
$\sin\chi$. A fit with $(1-\sin^2\chi)^{3/2}$ shows good agreement. 
The reason for this functional form with its non-analyticity at 
$\sin\chi=1$ is understandable from very general considerations:
\begin{figure}[t]
\centering{
\includegraphics[width=6cm]{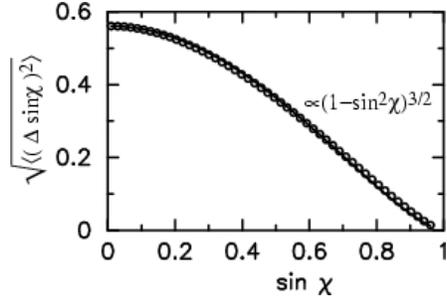}}
        \caption{RMS deviation of $\sin\chi$ from its initial 
        value for one iteration of the billiard map for the quadrupole 
        at $\epsilon=0.08$, averaged over final position ${\bar s}$. 
        Circles are from the exact ray dynamics, solid line is a fit 
        with $(1-\sin^2\chi)^{3/2}$.
        }
        \label{fig:kickaverage}
    \end{figure}

If one considers $F$ as a function of $\cos\chi$ instead of 
$\sin\chi$, then a Taylor expansion around $\cos\chi=0$ yields a 
mapping equation of the form
\be 
\cos{\bar\chi}\approx f_0({\bar s})+f_1({\bar s})\,\cos\chi 
+f_2({\bar s})\,\cos^2\chi+\ldots\label{analytex} 
\ee 
We expect $f_0\equiv 0$ 
since a trajectory starting with $\cos\chi=0$ must end up with 
$\cos{\bar\chi}=0$ (corresponding to ``rolling'' or ``grazing'' 
motion along the convex surface.  Using this expansion for 
$\sin\chi\to 1$ in Eq.\ (\ref{eq:sinchimap}), we obtain 
\be
\sin{\bar\chi}\approx\sin\chi-f_1({\bar s})\,f_2({\bar s})\, 
(1-\sin^2\chi)^{3/2}.\label{generalanalyt} 
\ee
The resulting $\sin\chi$ - dependence in Fig.\ \ref{fig:kickaverage} 
suggests that 
the map of the convex billiard should exhibit invariant 
curves in the ``whispering-gallery'' (WG) region ($\sin\chi\to 1$) at 
arbitrarily large deformations, 
where other KAM curves (such as the one belonging to the inverse 
golden mean winding number) have been broken up. 

The last statement is in fact implied by Lazutkin's 
theorem \cite{lazutkinbook}, 
which however requires 553 continuous derivatives of $r(\phi)$ to 
prove that invariant WG tori exist with nonzero Lebesgue measure. 
Here, we can go beyond the mere existence statement and ask what 
consequences the existence of a stable WG region has for the 
neighboring phase space. We shall find that only a 
much smaller number of 3 continuous derivatives enters the physical 
considerations describing the phaes space of the convex billiard. 
We base our argument on  an adiabatic approximation used in 
Ref.\ \cite{robnikberry}, which will be discussed further below. 
There, the unknown 
multiplier $f_1({\bar s})\,f_2({\bar s})$ of 
Eq.\ (\ref{generalanalyt}) is determined from geometric considerations 
to yield
\be 
\sin{\bar\chi}=\sin\chi-\frac{2\kappa'({\bar s})}{3\kappa^2({\bar s})}\, 
(1-\sin^2\chi)^{3/2},\label{sinusmap} 
\ee
where $\kappa(s)$ is the curvature and $\kappa'$ its derivative. 
In Ref.~\cite{robnikberry}, this together with a similar 
approximation for the position mapping function $G({\bar 
s},\sin\chi)$ is used to convert the amplitudes of the two mapping 
equations, Eq.\ (\ref{eq:sinchimap}) and Eq.\ (\ref{eq:smap}), 
into a differential equation:
\be\label{eq:differential}
\frac{d\sin\chi}{d s}\approx
\frac{\sin{\bar \chi}-\sin\chi}{{\bar s}-s}=\frac{F({\bar 
s},\sin\chi)}{G({\bar 
s},\sin\chi)}\approx\frac{F(s,\sin\chi)}{G({s,\sin\chi)}},
\ee
which can be solved by separation of variables to obtain an {\em 
adiabatic invariant curve}
\be\label{eq:adiacurve}
p(s)\approx
\sqrt{1-\left(1-\sigma^2\right)\,\kappa^{3/2}(s)}.
\ee
Here, we use the abbreviation 
\be 
p\equiv\sin\chi,
\ee 
which is the momentum conjugate to $s$. 
The intergation constant $\sigma$ parametrizes the value around which 
$p(s)=\sin\chi(s)$ oscillates. 

The range of validity of Eq.\ (\ref{sinusmap}) extends beyond the 
WG limit $\sin\chi\approx 1$, as Fig.\ \ref{fig:kickaverage} already 
suggests. A position mapping equation which also yields reasonable 
agreement for the whole range of possible initial $\sin\chi$ has been 
derived in Ref.\ \cite{thesis} by introducing a generating function 
$Z({\bar s},p)$ for the billiard map. From $Z$, the new momentum and 
old positions are obtained as partial derivatives, 
\be 
{\bar p} = \left.\frac{\partial Z}{\partial{\bar s}}\right|_p,\qquad 
s = \left.\frac{\partial Z}{\partial p}\right|_{\bar s}.\label{genrate} 
\ee 
This definition guarantees that the map is area preserving. The first 
equation above is just what we already obtained in Eq.\ (\ref{sinusmap}), 
so we can infer $Z({\bar s},p)$ by integrating Eq.\  (\ref{sinusmap})
over ${\bar s}$. This leads to 
\be\label{generatingfneqn} 
Z({\bar s},p) = p\,{\bar s}+\frac{2}{3\kappa({\bar s})}\,(1-p^2)^{3/2} 
+c(p), 
\ee 
where $c(p)$ is the integration constant which may still depend on 
$p$. Applying the second of  Eqs. (\ref{genrate}) to this result, we 
arrive at the position mapping equation,
\be 
s={\bar s}+c'(p)-\frac{2p}{\kappa({\bar s})}\,(1-p^2)^{1/2}. 
\label{positionmap} 
\ee 
This can in principle be inverted to get ${\bar s}$ as a function of 
$p$ and $s$. We dispose of the arbitrary $c'(p)$ in such a way that 
Eq.\ (\ref{positionmap}) reduces to the exact expression in the 
circular billiard where $\kappa\equiv 1$. The result is, reinstating 
$\sin\chi$ for $p$, 
\be
s={\bar s} 
-2\,\arccos(\sin\chi) +2\,(1-\frac{1}{\kappa({\bar s})})\, 
\sin\chi\,(1-\sin^2\chi)^{1/2}.  \label{effmapeqn2} 
\ee
In contrast to the analogous result in Ref.\ \cite{robnikberry}, this 
position map remains well-defined over the whole range of 
$\vert\sin\chi\vert=0\ldots 1$. The billiard shape 
enters in Eqs.\ (\ref{sinusmap}) and (\ref{effmapeqn2})
only through the curvature as a function of ${\bar s}$. 

The ``effective map'' 
as defined through Eqs.\ (\ref{sinusmap}) and (\ref{effmapeqn2}) 
reproduces the global structure as well as local detail of 
the true Poincar{\'e} sections for the Limacon and 
quadrupole billiards \cite{thesis}.
Although some additional rescaling of the deformation parameter 
$\epsilon$ is required for best agreement, one can use the 
effictive map to understand classical properties of the billiard, 
such as the existnce of Lazutkin's invariant tori. The utility of this 
approach consists in breaking the billiard problem up into two 
distinct subproblems: the geometric analysis leading to the 
Poincar{\'e} mapping on the one hand, and the nonlinear dynamics of 
that map on the other hand. We now wish to apply 
Eqs.\ (\ref{sinusmap}) and (\ref{effmapeqn2}) to our 
understanding of Lazutkin's theorem.

The adiabatic approximation leading to Eq.\ (\ref{sinusmap}) relies 
on the fact that for trajectories in the whispering-gallery region, 
there is a {\em separation of time scales} between slow 
changes in the average $\sin\chi$ and a fast circulation in arc 
length $s$ around the boundary; this separation becomes infinitely 
wide as $\sin\chi\to 1$, as required by Lazutkin's theorem. 
In other words, if multiple iterations of the 
map return ${\bar s}$ to an inifintesimal neighborhood $\Delta s$ of its 
initial value $s$, then the same will automatically be true for the 
second variable $\sin\chi$, in such a way that the derivative 
in Eq.\ (\ref{eq:differential}) exists. Now let us 
approach the adiabatic limit from the side of 
a chaotic trajectory described by the effective map, for which that 
derivative is ill-defined but a finite separation of time scales still 
exists. Then we can then ask for the local {\em diffusion constant}, 
defined as the proportionality constant between the rms spread of 
$\sin\chi$ (averaged over ${\bar s}$) and the number of 
mapping iterations $n$,
\bea
\langle\left(\Delta p(n)\right)^2\rangle &=&D(p_0)\,n\nonumber\\
D(p_{0})&=&\frac{4}{9}\,\label{diffusiveeqn}
\frac{1}{2\pi}\,(1-p_0^2)^3\,\int\limits_0^{L}
\left[\frac{{\dot\kappa}(s)}{\kappa^2(s)}\right]^2\,d s,
\eea
where $L$ is the circumference of the boundary.
The assumption of a diffusive growth in the variance of $\sin\chi$ 
leads us 
to make the {\em random phase approximation} \cite{lichtenberg} 
for $s$, as contained in the integral over $s$ above. 

In fact, this 
approximation is hard to justify in generic billiards with a mixed 
phase space, but improvements can in principle be made by including 
an average over a finite number of mapping steps in the definition of 
$D$. The main conclusion from Eq.\ (\ref{diffusiveeqn}) is that the 
diffusion constant follows the $\sin\chi$ dependence of 
Fig.\ \ref{fig:kickaverage} (squared) and hence vanishes for 
$\sin\chi\to 1$. Now define the 
(discrete) diffusion time in $p$, $\tau_p$ to be the number of 
iterations it takes
to diffuse across the whole allowed $p$-interval; define further 
the phase
randomization time $\tau_s$ as the number of reflections 
necessary before $s$
has wrapped once around the boundary. Then from Eq.\
(\ref{diffusiveeqn}), 
\be
\tau_p\propto\frac{1}{(1-p^2)^{3}},
\ee
and from Eq.\ (\ref{effmapeqn2}),
\be
\tau_{s}\propto\frac{1}{p\,\sqrt{1-p^2}}.
\ee
The proportionality constants are functions of the deformation alone. As
the whispering-gallery limit is approached, $\tau_p$ diverges much
faster than $\tau_{s}$, so that the existence of invariant curves can
be inferred as $\sin\chi\to 1$. Since
no more than the first derivative of $\kappa$ appears in the ``kick 
strength'' of the analytic mapping of Eqs.\ (\ref{sinusmap}) 
and (\ref{effmapeqn2}), this suggests that only three continuous 
derivatives of the boundary suffice to explain Lazutkin's tori.

\begin{figure}[t]
\centering{
\includegraphics[width=6cm]{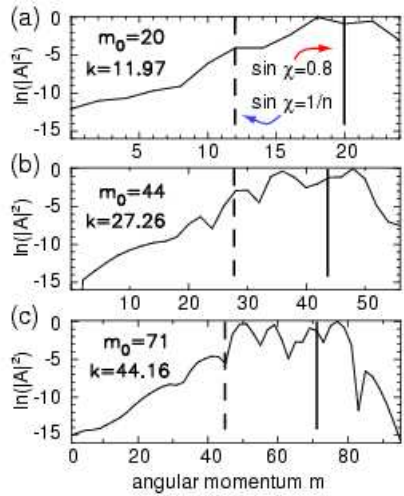}}
        \caption{
Decadic logarithm of the angular momentum coefficients $\vert 
A_{m}\vert^2$ in Eq.\ (\ref{psiintdecompgeneqn}) for three differnt 
quasibound states of the dielectric quadrupole with $n=2$ and 
$\epsilon=0.08$; all states are semiclassically quantized at 
approximately the same value of the adiabatic constant $\sigma$, at 
wavenumbers $kR\approx 11.97,\,27.26,$ and $44.16$. 
        }
        \label{fig:dynloc}
    \end{figure}
As an example of how the above remarks on phase space transport 
properties in a convex billiard help us understand the quasibound 
states of the corresponding resonator, Fig.\ \ref{fig:dynloc} shows 
how {\em dynamical localization} can be discerned in the numerical 
solutions of the wave problem. 
The same states displayed here have been investigated in 
Ref.\ \cite{nature} as a function of the deformation $\epsilon$. 
There, an adiabatic quantization based on the invariant curves
Eq.\ (\ref{eq:adiacurve}) was introduced, according to which all three 
states were found to correspond to approximately the same value of 
$\sigma$, i.e. the adiabatic invariant characterizing the phase-space 
location of the state. The wavenumbers $k$ of the lowest and highest 
state in Fig.\ \ref{fig:dynloc} differ by roughly a factor of 
four, and as a consequence the decay 
rates in the limit of a circular resonator range over approximately 
ten orders of magnitude. However, at large deformation $\epsilon$ 
where the escape from the resonator is dominated by classical ray 
diffusion, the resonance widths $\delta k$ are found to be nearly 
wavelength independent. As a correction to this classical 
behavior, one observes a tendency toward slightly faster decay at 
larger wavenumber $k$. This correction, together 
with the fact that the adiabatic quantization for the resonance 
positions actually agreed better with the exact results at smaller $k$ 
led to the hypothesis \cite{nature} that dynamical localization is present 
in the states under consideration, especially at low $k$. 

Figure \ref{fig:dynloc} allows us to identify qualitatively the effect 
of dynamical localization in a mixed phase space with open boundary 
conditions. The solid vertical lines in the figure indicate the 
semiclassically expected maxima of the angular momentum distribution 
for the three wavefunctions. This semiclassical quantization is 
performed by applying the EBK method to the invariant curves of 
Eq.\ (\ref{eq:adiacurve}). The quantized values of $\sigma$ are then 
translated back to angular momentum by using the approximate 
semiclassical relation known from the circle,
\be\label{eq:sinchim}
m=n\,kR\,\sin\chi.
\ee
We use this together with our 
classical considerations to estimate the spread in $m$ that is evident 
in the wavefunctions. 

The interval of $m$ on the horizontal axis corresponds to the 
range $0\le \sin\chi\le 1$; the dashed vertical lines indicate the 
``ionization border'' $\sin\chi=1/n$. 
At $\epsilon=0.08$ in the quadrupole, unbroken KAM curves 
exist only above $\sin\chi\approx 0.9$. The sharp 
falloff in $|A_{m}|$ at large $m$ is due to the classical 
inaccessibility of the 
high-$\sin\chi$ region by diffusion -- it directly shows the 
imbalance in the nonlinearity between large and small $\sin\chi$. 

Exponential localization is identifiable in Fig.\ \ref{fig:dynloc} 
only to the left of a {\em plateau} surrounding the semiclassical 
maxima, of width $\Delta m\approx 4$ in (a) and $\Delta m\approx 10$ 
in (b). The explanation for this is that the adiabatic 
curve, Eq.\ (\ref{eq:adiacurve}), 
for $\sigma=0.8$ oscillates between $\sin\chi_{max}\equiv
=0.89$ and $\sin\chi_{min}=0.72$.
The latter translates to an angular momentum spread of 
$\Delta m\approx 4,\,9,\,$ and $14$, respectively, for the states 
quantized at $kR\approx 11.97,\,27.26,$ and $44.16$. The agreement 
with the Figure confirms that the 
semiclassical maximum, ionization border and width of the plateau in 
$m$ all scale proportional to $kR$. This leaves us with an $m$ 
interval $\delta m$ to the right of the escape threshold, of width 
$\delta m\approx n\,kR\,(\sin\chi_{min}-\sin\chi_{c})=n\,kR\,(0.72 - 
0.5)$, corresponding to a classically diffusive region in which a 
probability decay should be observed. In the Figure, 
an exponential decrease away from the semiclassical plateau is  
discernible for $kR=11.97$ (a) and $27.26$ (b), whereas the state at
$kR=44.16$ (c) 
exhibits large angular momentum components over the whole classically 
confined region. The localization lengths $\xi$ estimated from the observed 
slopes for the two lower-$kR$ states are approximately in a {\em 
ratio} of $2\,:\,1$, in reasonable agreement with the factor of two 
between the respective wavelengths. This is the expected 
behavior \cite{borgonovi}, 
because the diffusion constant entering $\xi$ is the same in all 
resonances. 

Angular momentum as a prefered basis for measuring 
dynamical localization is a useful initial choice in the oval 
billiard, but as the existence of the plataeus above indicates, it is 
not strictly the correct one. Recall that $\sigma$, and not 
$\sin\chi$ or angular momentum, is the adiabatic invariant in the 
semiclassical quantization for the states in Fig.\ \ref{fig:dynloc}. 
This can be made very clear by comparing to the special case of 
the ellipse where the adiabatic curve Eq.\ (\ref{eq:adiacurve}) 
becomes exact. Then $\sigma$, which in the circle is the angular 
momentum, acts as a constant of the motion while $\sin\chi$ still 
oscillates. Clearly, there is no diffusion although the angular 
momentum decomposition shows a spread $\Delta m$ whose width is 
determined by the eccentricity. For the oval billiard, this means 
that $\sigma$ should be considered as the diffusing variable. 
Classically, the transformation from $\sin\chi$ to $\sigma$ is given 
by Eq.\ (\ref{eq:adiacurve}). Wave-mechanically, the goal will be to 
project the true wave function $\psi$ onto a Treftz basis defined by 
these adiabatic invariant curves. 

Judging by the results presented here, this is a worthwile program for 
future work because a better-adapted basis significantly expands the 
interval over which one is allowed to assume the separation of time 
scales which leads to classical diffusion in the first place. 
In billiards that remain close to a circle, such as 
a short stadium \cite{borgonovi} or rough billiard 
\cite{frahm,starykh} this problem does not arise. 
However, in oval billiards, these classical considerations apply. 
Moreover, the breakdown of the Rayleigh hypothesis makes it not only 
desirable but necessary to abandon angular momentum as the basis in 
which to detect dynamical localization. 

I would like to thank Steve Tomsovic for valuable discussions.

%%%%%%%%%%%%%%%%%%%%%%%%%%%%%%%%%%%%%%%%%%%%%%%%%%%%%%%%%%%%%%%%%%%%%%%%%%
%
%        Figure captions included here:
%
%%%%%%%%%%%%%%%%%%%%%%%%%%%%%%%%%%%%%%%%%%%%%%%%%%%%%%%%%%%%%%%%%%%%%%%%%%
    
\end{document}
\end